# Novel boron nitride polymorphs with graphite-diamond hybrid structure

Kun Luo(罗坤)[1†], Baozhong Li(李宝忠)[1†], Lei Sun(孙磊)[1], Yingju Wu(武英举)[1,2*], Yanfeng Ge(盖彦峰)[1,2], Bing Liu(刘兵)[1], Julong He(何巨龙)[1], Bo Xu(徐波)[1], Zhisheng Zhao(赵智胜)[1*], Yongjun Tian(田永君)[1]

[1]Center for High Pressure Science (CHiPS), State Key Laboratory of Metastable Materials Science and Technology, Yanshan University, Qinhuangdao 066004, China

[2]Key Laboratory for Microstructural Material Physics of Hebei Province, School of Science, Yanshan University, Qinhuangdao 066004, China

†These authors contributed equally to this work

*Corresponding authors. Email: zzhao@ysu.edu.cn; wuyj@ysu.edu.cn

**Abstract:** Both boron nitride (BN) and carbon (C) have sp, $sp^2$ and $sp^3$ hybridization modes, and thus resulting in a variety of BN and C polymorphs with similar structures, such as hexagonal BN (hBN) and graphite, cubic BN (cBN) and diamond. Here, five types of BN polymorph structures were proposed theoretically, inspired by the graphite-diamond hybrid structures discovered in recent experiment. These BN polymorphs with graphite-diamond hybrid structures possessed excellent mechanical properties with combined high hardness and high ductility, and also exhibited various electronic properties such as semi-conductivity, semi-metallicity, and even one- and two-dimensional conductivity, differing from known insulators hBN and cBN. The simulated diffraction patterns of these BN hybrid structures could account for the unsolved diffraction patterns of intermediate products composed of "compressed hBN" and diamond-like BN, caused by phase transitions in previous experiments. Thus, this work provides a theoretical basis for the presence of these types of hybrid materials during phase transitions between graphite-like and diamond-like BN polymorphs.

Boron nitride (BN), known as the twin brother of carbon (C), has many similarities to C, such as the hybridization modes of sp, sp$^2$, and sp$^3$, and various polymorphs.[1-11] There are four known experimental BN polymorphs, including hexagonal BN (hBN),[1,2] rhombohedral BN (rBN),[4,5] wurtzite BN (wBN),[6] and cubic BN (cBN).[8] Among these, hBN and rBN have graphite-like layered structures, where the B and N atoms are alternately connected by sp$^2$ bonds to form a honeycomb layered structure. These layers are then stacked with AaAa and ABCABC sequences, respectively.[1,2,4,5] Both wBN and cBN have diamond-like structures, that correspond to hexagonal diamond (HD) and cubic diamond (CD), respectively.[6,8] The wBN and cBN structures can be considered to be constructed by sp$^3$ bonding of the B and N atoms in the adjacent hBN or rBN layers, and the stacking sequences of the puckered layers in structures are also AaAa and ABCABC, respectively.[12] These known polymorphs exhibit distinct physical properties. For example, hBN is a soft insulator with a bandgap of 5.9 eV,[1,13] while cBN is a superhard insulator with a bandgap of 6.3 eV.[14]

Over the years, the structural transformations between graphite-like BN and diamond-like BN polymorphs have been studied a lot.[6-8,15-22] Theoretically, hBN can transform into wBN by directly compressing and puckering the basal planes, owing to the same stacking sequences of AaAa.[16] In early experimental studies, under the conditions of high pressure and relatively low temperatures (P> ~8.5 GPa, T< ~1600 °C), hBN tends to transform into wBN with structure only identified by X-ray diffraction (XRD) technology, and the higher the pressure, the lower the temperature required for phase transition.[7,17] Until recently, millimeter-sized wBN blocks stabilized by three-dimensional networks of planar defects were synthesized using hBN single crystals as the precursor, and the corresponding structure was confirmed by atomic-resolution high-angle annular dark-field (HAADF) images.[6] In addition, previous experimental studies have shown that hBN tends to transform into cBN rather than wBN under pressure when the synthesis temperature was higher than 1600 °C.[7,17,19] Also, the transformation from wBN to cBN was observed when the shock-formed wBN powder was treated at high pressure and high temperature.[17] The formation of cBN was possibly due to its higher thermodynamic stability at high temperature under pressure.[17] For rBN, it can transform into cBN, owing to the same ABCABC stacking sequences.[20,21] However, the layers of rBN slip easily,[3,23] and a large number of stacking faults and irregular stackings will form under low pressure,[18,23] resulting in the formation of wBN in local areas during transition.[18]

In the above experimental studies, a diffraction peak corresponding to the interlayer spacing of ~3.1 Å, which was smaller than that of hBN, often appeared in the incompletely transformed samples from graphite-like BN to diamond-like BN.[17,18,24-26] For a long time, this was simply assigned to the compressed layered BN and called "compressed hBN",[17,18,24,25] similar to the "compressed graphite" observed in partially transformed graphite samples.[27,28] In recent years, "compressed hBN" has been studied by high-resolution transmission electron microscopy (HRTEM),[26] and the results showed that the layered BN was folded and bent due to the limitations of the adjacent cBN grains, resulting in a reduction in interlayer spacing to ~3.06-3.34 Å.[26] In addition, slight slip would occur between the BN layers due to bending and folding, and as a result, the symmetry of the layered structure was no longer hexagonal and morphed into monoclinic in some local areas.[26] Until recently, it was generally believed that the incompletely transformed BN samples were a mixture of "compressed hBN" and diamond-like BN, and the formation of the compressed layered BN was caused by residual stress.[17,18,24-26]

Recently, two types of graphite-diamond hybrid structures have been proposed in natural and laboratory-shocked diamond-related samples by HRTEM observations, one of which consisted of few-layered graphene contiguous with the Pandey (2×1) reconstructed surfaces of diamond {111} planes separated/interfaced by weak van der Waals interactions, and the other consisted of graphitic layers inserted within the Pandey (2×1) reconstructed surfaces of {113} diamond with a semi-coherent interface structure.[29,30] Furthermore, four other types of graphite-diamond hybrid structures with completely coherent interface structures were confirmed through atomic-resolution HAADF images of partially transformed graphite samples recovered from static compression.[31] During the formation of these hybrid structures,[31] compression-induced bending of the graphite layers resulted in inconsistencies between the interlayer distances and stacking orders at different local areas of the structure. Furthermore, diamond could form by interlayer bonding in local areas with reduced interlayer spacing and an appropriate stacking order; however, other areas still maintained graphite structure.[31] As a result, the graphite domains were intimately connected with the diamond domains through a completely coherent interface in the hybrid structure.[31] The graphite component in the hybrid structure had an interlayer spacing of ~3.1 Å, and this could explain the origin of "compressed graphite" observed in the experiment.[27,28,31]

Considering the phase transition similarities between BN and C, in this study, five types of BN hybrid structures composed of periodic graphite-like BN and diamond-like BN units were constructed. Four types were constructed on the basis of graphite-diamond hybrid structures,[31-33] and the fifth structure is newly proposed by redesigning the orientation relationship of the graphite-like BN and wBN units. The five types of BN hybrid structures were thermodynamically more stable than hBN under pressure, and could even directly transform into cBN or wBN at high pressure. These hybrid structures had a theoretical hardness of 19-31 GPa, which was higher than that of $Al_2O_3$ and comparable to that of SiC. They also had a variety of peculiar electronic properties varying from semi-conductivity, semi-metallicity, and then to metallicity.

The BN hybrid structures were constructed with the Materials Visualizer module in Materials Studio (Accelrys Software Inc.).[34] The calculations were performed based on density functional theory (DFT) implemented by CASTEP code,[35] and ultrasoft pseudopotentials were used.[36,37] The local density approximation (LDA) with the Ceperley-Alder exchange-correlation potential parameterized by Perdew and Zunger (CA-PZ) was used to structural optimization and to calculate the total energies, elastic properties and phonon spectra.[38,39] A $k$-point sampling[40] of $2\pi \times 0.03$ Å$^{-1}$ and a plane-wave cutoff of 600 eV were used. The selected calculation parameters were all tested to ensure that the energy convergence was less than 1 meV per atom.[41,42] Structural relaxation was stopped when the total energy change, the maximum displacement, stress, and Hellmann–Feynman force were less than $5.0 \times 10^{-6}$ eV/atom, $5.0 \times 10^{-4}$ Å, 0.02 GPa, and 0.01 eV/Å, respectively. The symmetry $k$-path in the Brillouin zone of the primitive cells was determined by the SeeK-path,[43] which was used to calculate the phonon spectra and band structures. The phonon spectra were calculated via the finite displacement method.[44] To validate our computational scheme, benchmark calculations were carried out for cBN, and the calculated lattice parameter of $a = 3.58$ Å was in good agreement with the experimental results of $a = 3.62$ Å for cBN.[45] The calculated bulk modulus of cBN was 389 GPa, which was in agreement with the experimental value of 395 GPa.[45] We also constructed the BN hybrid structures with nanoscale graphite-like and diamond-like units, and simulated their XRD patterns with the Reflex module in Materials Studio,[34] to compare with the experimental diffraction patterns,[17,18,24] where the wavelength of the selected X-ray was 1.54 Å.

The five BN hybrid structures are shown in Fig. 1, and the corresponding connection modes between the graphite-like and diamond-like BN units in the structures are also highlighted in the figure. Four of these were constructed based on graphite-diamond hybrid structures[31-33] previously named Gradia[33] (Fig. 1a-d), and the fifth was a newly proposed structure (Fig. 1e).[46] The first two hybrid structures in Fig. 1(a) and (b) were composed of graphite-like BN and cBN units (analogous to CD) with an orientation relationship of $[010]_{hBN}//[1\bar{1}0]_{cBN}$. When viewed along $[1\bar{1}0]_{cBN}$, it can be seen that the $(111)_{cBN}$ and $(11\bar{1})_{cBN}$ planes formed a rhombic pattern with a side-length of 2.21 Å, and the graphite-like BN units were connected to the vertices with either obtuse or acute angle of the rhombic pattern in cBN. These two BN hybrid structures were referred to as Gradia-CO and Gradia-CA in BN forms,[46] respectively, in analogy to the graphite-diamond hybrid structures.[31-33] The third and fourth hybrid structures were composed of graphite-like BN and wBN units (analogous to HD) with an orientation relationship of $[010]_{hBN}//[010]_{wBN}$ (Fig. 1c and d). When viewed along $[010]_{wBN}$, the $(100)_{wBN}$ and $(002)_{wBN}$ planes formed a 2.21×2.11 Å rectangular pattern. During compression, the adjacent BN layers could either partially buckle into a boat configuration and transform into $(100)_{wBN}$ with a $d$-spacing of 2.21 Å, or partially pucker into a chair configuration and transform into $(002)_{wBN}$ with a $d$-spacing of 2.11Å. These two hybrid structures are referred to as Gradia-HB and Gradia-HC in BN forms,[46] respectively, in analogy to those in carbon.[31,33]

Furthermore, a new BN hybrid structure was proposed by redesigning the orientation relationship between the graphite-like and wBN units,[46] as shown in Fig. 1(e). The orientation relationship newly constructed structure was $[210]_{hBN}//[001]_{wBN}$, which was different from that in Gradia-HB and Gradia-HC. When viewed along $[001]_{wBN}$, the adjacent BN layers could partially pucker into a zigzag configuration and transform into $(100)_{wBN}$, therefore, it was referred to as Gradia-HZ in BN form.[46]

The space group, lattice parameters and atomic Wyckoff positions of the representative hypothetical crystal structures after optimization at ambient pressure are shown in Table S1. The interlayer spacing of the graphite-like BN units in these hybrid structures ranged from 2.87-3.36

Å, which was consistent with that of "compressed hBN" previously observed in the HRTEM images.[26] In addition, the detailed structure information of Gradia-HZ in carbon form is also listed in Table S1 and shown in Fig. S2(a) for future experimental verification.

The formation enthalpies of the BN hybrid structures with respect to hBN as a function of pressure are shown in Fig. 2(a). Typically, these hybrid structures would become thermodynamically more stable under pressure, and would easily transform into diamond-like BN above a certain pressure. Among these, the enthalpy of the Gradia-CA structure was lower than that of hBN above a pressure of 19 GPa. Under pressure of 2-9 GPa, the Gradia-CO structure directly transformed into cBN, and the Gradia-HB, Gradia-HC and Gradia-HZ structures directly transformed into wBN, indicating the low transition energy barriers of the hybrid structures to cBN or wBN. Fig. 2(b) shows the transformation process of the Gradia-HZ structure to wBN during optimization at a pressure of 9 GPa. This indicated a gradual transformation process from graphite-like BN units to wBN units through linkage of the graphite-like BN layers at the junction in the hybrid structure. Finally, the hybrid structure completely transformed into wBN.

To assess the dynamic stabilities of the BN hybrid structures, we calculated their phonon spectra at ambient pressure, as shown in Fig. S1. No imaginary frequency was observed for the hybrid structures in the entire Brillion zone, confirming that they were all dynamically stable. We also calculated their ambient-pressure elastic constants, which are listed in Table S2. The calculated 13 independent elastic constants of the BN hybrid structures satisfied the Born stability criterion for monoclinic structures,[47-50] demonstrating they were mechanically stable. In addition, the Gradia-HZ structure in carbon form was also predicted to be mechanically and dynamically stable at ambient pressure (Table S2 and Fig. S2c).

The mechanical properties of the hybrid structures are listed in Table 1. The bulk moduli ($B$) and shear moduli ($G$) for the five BN hybrid structures ranged from 177-282 and 126-198 GPa, respectively. By contrast, the $B$ and $G$ of cBN were calculated as 389 and 404 GPa, respectively. The BN hybrid structures had high $B/G$ values of 1.2-1.7, implying that their ductility was better than cBN (and wBN) with a $B/G$ of ~0.96. On the basis of the hardness model,[51,52] the Vickers hardness values of the five BN hybrid structures were estimated to be 19-31 GPa, which was higher than that of $Al_2O_3$ and comparable to that of SiC. The Gradia-HZ structures in both carbon and BN forms had similar mechanical properties. The *B*, *G* and Vickers hardness values of the Gradia-HZ structure in carbon form were predicted to be 305 GPa, 201 GPa, and 25 GPa, respectively.

The electronic properties of the BN hybrid structures were also investigated, and their band structures are shown in Fig. 3. Obviously, both Gradia-CO and Gradia-HB had a valence band and conduction band across the Fermi level, indicating that they were metallic (Fig. 3a and c). Gradia-CA and Gradia-HC were semi-metallic due to only an extremely small portion of valence bands across the Fermi level (Fig. 3b and d). The Gradia-HZ structures in both BN and carbon forms were semiconductors with bandgaps of 2.9 and 0.26 eV, respectively (Fig. 3e and Fig. S2d). Because LDA generally underestimates ~30-40% of bandgaps,[53] the band gaps of the Gradia-HZ structures could be larger. It is worth noting that the Gradia-HZ structure in carbon form was a narrow bandgap semiconductor with promising photoelectric applications.

To reveal the unique conductive behaviors of the metallic Gradia-CO and Gradia-HB structures in BN form, the electron orbits originating from the bands across the Fermi level were calculated and are shown in Fig. 4. We found that the bands across the Fermi level primarily came from the atoms at the junction regions between the graphite-like and diamond-like BN units in the hybrid structures. This was because these atoms at the junction had nonplanar $sp^2$ bonding and distorted tetrahedral $sp^3$ bonding, which were deviated from the equilibrium state and led to unexpected electronic properties. Eight nonequivalent atoms were observed at the junctions in the Gradia-CO and Gradia-HB structures, which were labeled as B1, B2, B3, N1, N2, N3, N4, and N5, respectively. The partial density of states (PDOS) of these atoms at the junction were specifically calculated and are shown in Fig. 4(a) and (b). Combined with the electron orbits and PDOS analysis, we found that the metallicity of the Gradia-CO structure was mainly due to the B1, B2, N1, N2, and N3 atoms at the junction. The electron orbits of these atoms were next to each other, forming a conductive network on the YZ plane, which was divided in the X-axis direction, causing the Gradia-CO to exhibit two-dimensional planar conductivity. However, the metallicity of Gradia-HB was mainly due to the B1, B3, N1, N2, and N4 atoms at the junction. The corresponding electron orbits of these atoms formed a conductive channel along the Z-axis direction, but they were separated in both the Y- and X-axis directions, causing Gradia-HB to only exhibit one-dimensional linear conductivity. More novel electronic properties were expected to be obtained by changing the sizes of the graphite-like and diamond-like units in the BN hybrid structures.[32]

Finally, the presence of these BN hybrid structures was further demonstrated by comparing the diffraction patterns with the samples composed of mixed "compressed hBN" and diamond-like BN obtained in previous experiments.[17,18,24] Considering that the graphite-like and diamond-like components in the experimental samples should had nanometer or even micrometer scale dimensions, we constructed the BN hybrid structures with nano-sized graphite-like and diamond-like units, and then simulated their diffraction patterns, as shown in Fig. 5. The sizes of the graphite-like and diamond-like units currently constructed in the five hybrid structures were 6.8 and 21.9 nm for Gradia-CO, 7.2 and 39.6 nm for Gradia-CA, 6.1 and 32.6 nm for Gradia-HB, 2.3 and 20.4 nm for Gradia-HC, and 5.4 and 21.6 nm for Gradia-HZ, respectively. In general, the simulated XRD patterns of the Gradia-CO and Gradia-CA structures can explain the experimental diffraction patterns of the intermediate products containing "compressed hBN" and cBN recovered after compressing polycrystalline hBN at a pressure of 7.7 GPa and temperatures of 2000-2300 °C.[24] The simulated XRD patterns of the Gradia-HB, Gradia-HC and Gradia-HZ structures were regarded as the superpositions of the diffraction patterns of wBN and graphite-like BN with reduced interlayer spacing; thus, they could account for the experimental diffraction patterns of the samples containing "compressed hBN" and wBN.[17,18] It should be noted that changing the nanometer size of the graphite-like and diamond-like structural units in the BN hybrid structure would not fundamentally change the simulated diffraction patterns. Therefore, these BN hybrid structures could be considered as intermediate polymorphs between graphite-like and diamond-like BN, which may have been obtained in previous experiments. In the near future, the atomic resolution images of these BN hybrid structures were expected to be obtained through state-of-the-art HAADF investigations to finally confirm these hybrid structures.

In summary, five types of BN hybrid structures composed of graphite-like and diamond-like BN units were proposed based on graphite-diamond hybrid structures discovered in recent experiment,[31,33] and studied by the first principles calculations. The investigated five hybrid structures were mechanically and dynamically stable at ambient pressure, and four of these structures could directly transform into cBN or wBN under high-pressure conditions. These hybrid structures exhibited excellent mechanical properties including high hardness and high ductility, and various electronic properties ranging from semi-conductivity, semi-metallicity, and even to one- and two-dimensional conductivity. The simulated diffraction patterns of these hybrid structures could explain the unsolved diffraction patterns of previously recovered samples containing "compressed hBN" and diamond-like BN.[17,18,24] Thus, these BN hybrid structures could be intermediate polymorphs[46] during the transformation between graphite-like and diamond-like BN. As a result, this work could greatly promote the extensive experimental exploration of BN hybrid materials with unique physical properties.

*Acknowledgments*. This work was supported by National Natural Science Foundation of China (Nos. 52090020, 91963203, U20A20238, 51772260, 52073245, 51722209), the National Key R&D Program of China (2018YFA0703400 and 2018YFA0305900), the Natural Science Foundation for Distinguished Young Scholars of Hebei Province of China (E2018203349), and the Talent research project in Hebei Province (2020HBQZYC003).

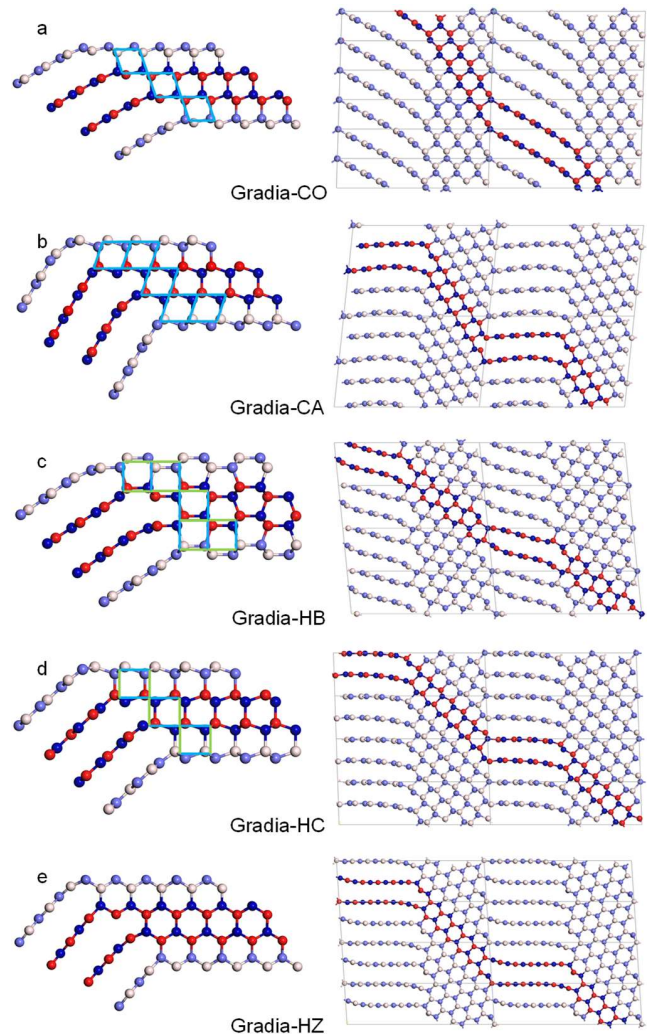

**Fig. 1.** Five types of BN hybrid structures composed of graphite-like and diamond-like BN units. (a-d) Gradia-CO, Gradia-CA, Gradia-HB, and Gradia-HC structures in BN forms constructed from the graphite-diamond hybrid structures.[31-33] (e) A new hybrid structure, Gradia-HZ, was constructed by changing the orientation relationship between the graphite-like BN and wBN units, where the left of each figure represents the corresponding connection modes of the graphite-like and diamond-like BN units at the junction. In the Gradia-CO and Gradia-CA structures, the graphite-like BN layers were connected to the vertices with either an obtuse or an acute angle in the blue-colored rhombus pattern in the cBN unit, respectively. In the Gradia-HB and Gradia-HC structures, the graphite-like BN layers were connected to the vertices of the green short or blue long sides of the rectangular pattern in the wBN unit, respectively. In the Gradia-HZ structure, the graphite-like BN layers were partially linked in a zigzag configuration, resulting in the formation of wBN unit. The right of each figure shows the corresponding periodic crystal structure after optimization at ambient pressure, and the interlayer spacings of the graphite-like BN units in the five hybrid structures were in the range of 2.87-3.36 Å. The bilayer highlighted in the BN hybrid structures corresponds to the adjacent layers of the original graphite-like BN before transformation. In the structure, boron atoms are marked as pink and red, and nitrogen atoms are marked as light blue and dark blue.

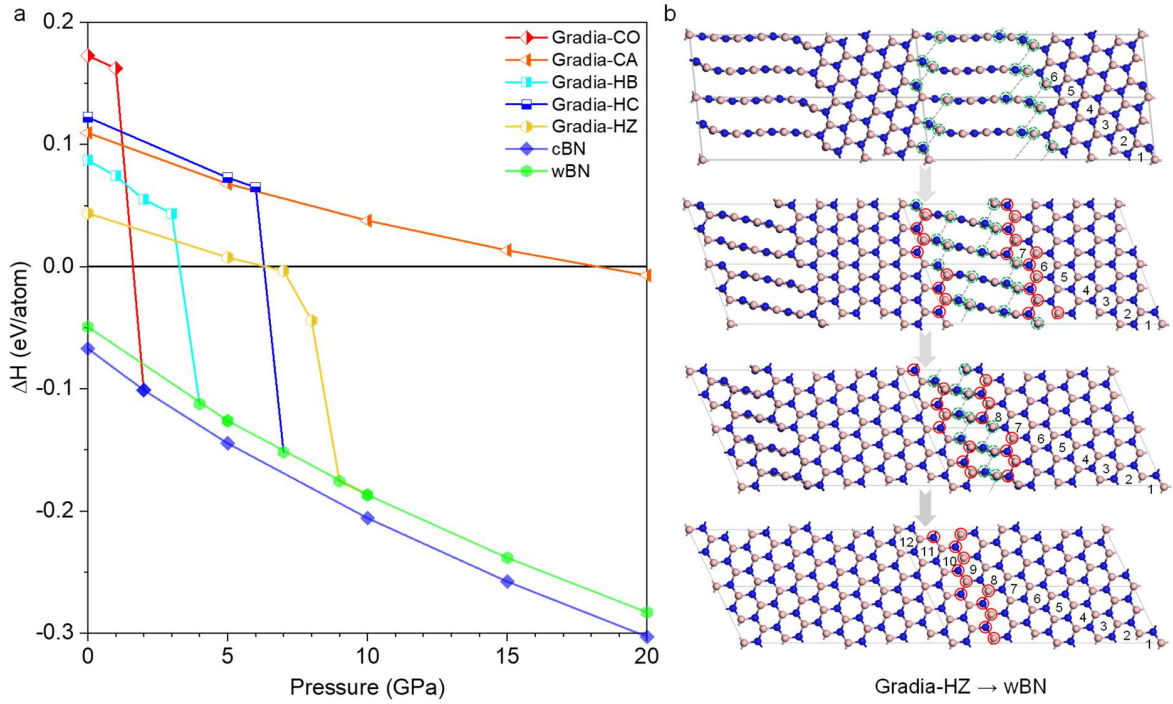

**Fig. 2.** Enthalpy changes and phase transitions of the BN hybrid structures under pressure. (a) Changes in enthalpy of the BN hybrid structures with respect to hBN as a function of pressure. The five hybrid structures were metastable at ambient pressure. The Gradia-CA structure was more stable than hBN above 19 GPa, and the other four hybrid structures directly transformed into cBN or wBN during optimization at pressures less than 9 GPa, indicating the low transition energy barrier from the four hybrid structures to cBN or wBN. (b) Full structure optimization of Gradia-HZ at 9 GPa, showing the gradual transformation to wBN. The circles with green dashed lines and the red solid lines in the structure represent the atoms that are about to undergo sp$^3$ bonding in the adjacent layer and the atoms that have completed sp$^3$ bonding, respectively.

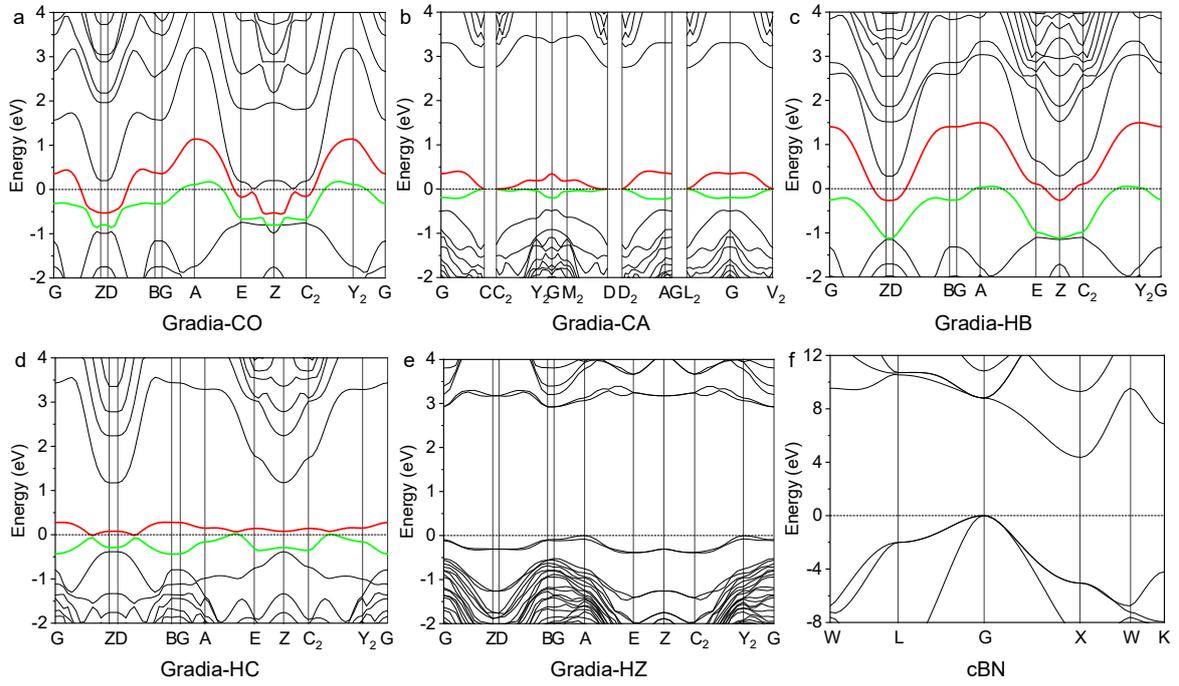

**Fig. 3.** Versatile electronic properties of the BN hybrid structures. Band structures of (a) Gradia-CO, (b) Gradia-CA, (c) Gradia-HB, (d) Gradia-HC, (e) Gradia-HZ, and (f) cBN calculated at ambient pressure. Gradia-CO and Gradia-HB were metallic, Gradia-CA and Gradia-HC were semi-metallic, and Gradia-HZ was a semiconductor with a bandgap of 2.9 eV, which was smaller than that of cBN (4.4 eV). For the metallic and semi-metallic hybrid structures, the bands near the Fermi level are marked by with red and green colors.

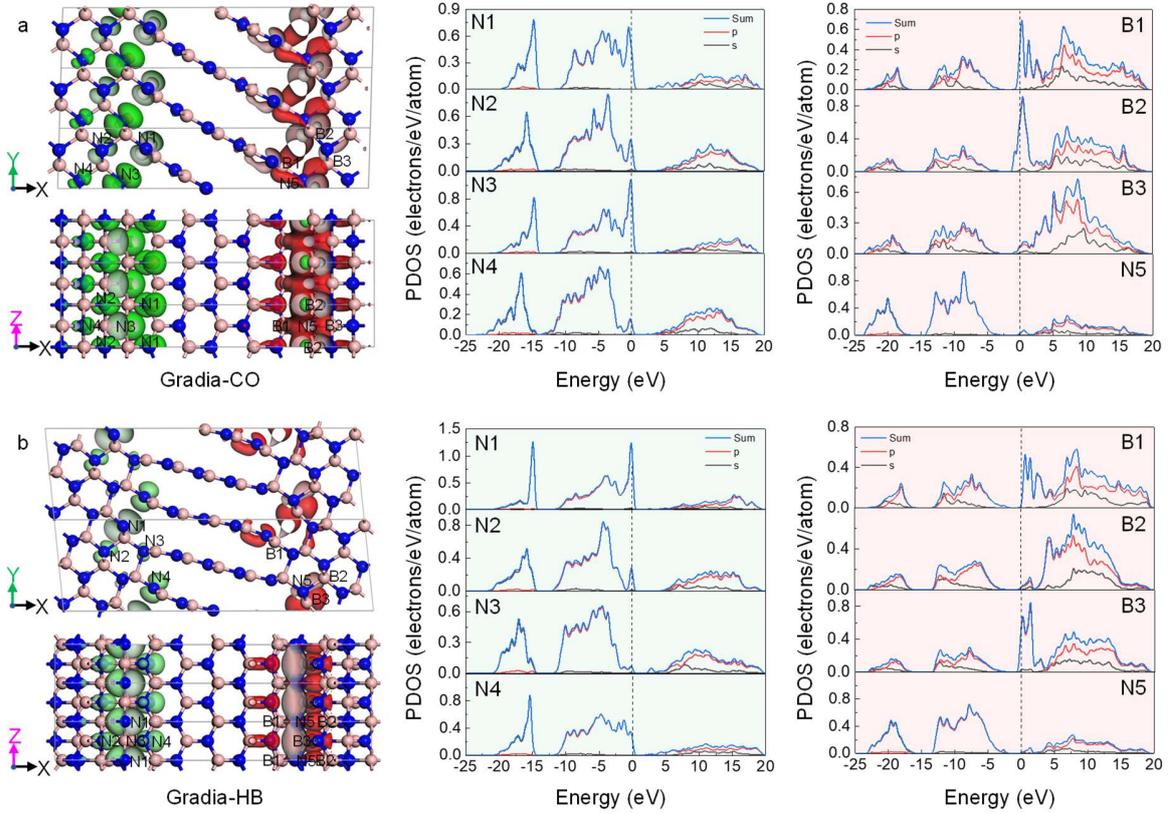

**Fig. 4.** Peculiar conductive behavior of the metallic BN hybrid structures. (a and b) Electron orbits and partial density of states (PDOS) of the Gradia-CO and Gradia-HB structures, respectively. The electron orbits originating from the red and green bands across the Fermi level in Fig. 3(a) and (c) are shown in the images, which were mainly come from the atoms at the junction between the graphite-like and diamond-like BN units in structures. The electron orbits were displayed by selecting electron density isosurface of 0.04 e/Å$^3$, and the PDOS of the corresponding labeled B and N atoms at the junction are also shown. The metallicity of Gradia-CO was mainly contributed by the B1, B2, N1, N2, and N3 atoms at the junction, and the electron orbits of these atoms formed a conductive network on the YZ plane, which was divided in the X-axis direction, resulting in two-dimensional planar conductivity of the structure. The metallicity in Gradia-HB was mainly from the B1, B3, N1, N2, and N4 atoms at the junction, and the electron orbits of these atoms formed a conductive channel along the Z-axis direction. However, they were separated in both the Y- and X-axis directions, causing Gradia-HB to exhibit one-dimensional linear conductivity.

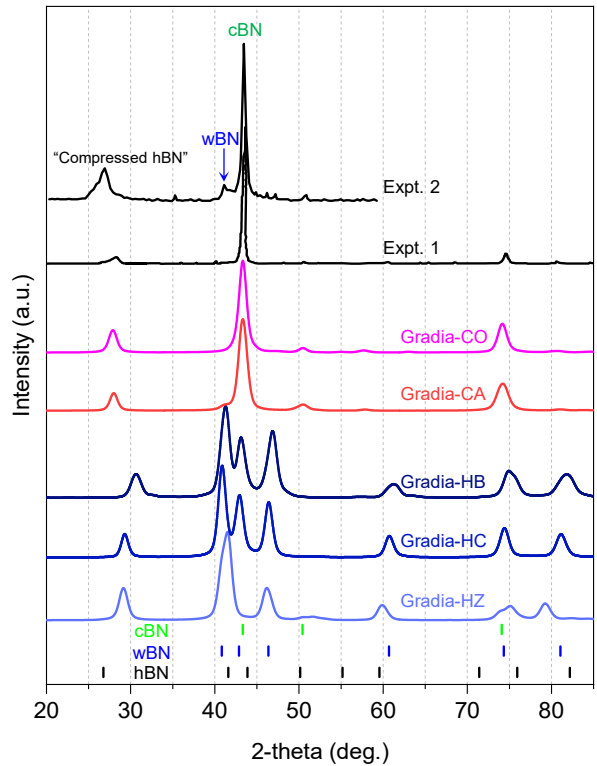

**Fig. 5.** Simulated XRD patterns of the BN hybrid structures, compared to the experimental patterns.[17,18,24] Expt. 1 shows the experimental diffraction pattern of the sample transformed from hBN treated at a pressure of 7.7 GPa and temperatures of 2000-2300 °C,[24] and Expt. 2 shows that of the sample obtained by treating rBN at 10 GPa and 1200 °C.[18] The simulated XRD pattern of the BN hybrid structure shows the diffraction characteristics of both graphite-like and diamond-like BN, and the interlayer spacing of the graphite-like units in the hybrid structure was smaller than that of hBN, which would vary according to the specific structure type. The simulated XRD patterns can account for the experimental diffraction patterns of the samples containing the so-called "compressed hBN" and cBN or wBN.[17,18,24] Thus, the BN hybrid structures could be considered as intermediate polymorphs between the graphite-like and diamond-like BN.

**Table 1.** Calculated ambient-pressure bulk moduli ($B$, GPa), shear moduli ($G$, GPa), $B/G$, Young's moduli ($E$, GPa), and Vickers hardness values ($H_v$, GPa) of the five BN hybrid structures as well as cBN and wBN.

| Structure | $B$ | $G$ | $E$ | $B/G$ | $H_v$ |
|---|---|---|---|---|---|
| cBN | 389 | 404 | 902 | 0.96 | 67.3 |
| wBN | 390 | 405 | 901 | 0.96 | 67.4 |
| Gradia-CO | 177 | 126 | 306 | 1.4 | 19.2 |
| Gradia-CA | 267 | 177 | 435 | 1.5 | 22.5 |
| Gradia-HB | 241 | 198 | 465 | 1.2 | 30.9 |
| Gradia-HC | 282 | 166 | 416 | 1.7 | 18.7 |
| Gradia-HZ | 247 | 152 | 378 | 1.6 | 18.5 |